\documentclass[twocolumn]{aastex63}

\usepackage{graphics,graphicx}
\usepackage{times}

\usepackage[english]{babel} 
\usepackage[T1]{fontenc} 
\usepackage{lmodern} 
\usepackage{float} 
\usepackage{amsmath} 
\usepackage{amssymb} 

\usepackage[autostyle]{csquotes}

\usepackage{calrsfs,euscript,mathrsfs,amssymb}
\usepackage{gensymb}
\usepackage[T5,T1]{fontenc}

\begin{document}

\title{Probing the connection between IceCube neutrinos and MOJAVE AGN}

\affiliation{III. Physikalisches Institut, RWTH Aachen University, D-52056 Aachen, Germany}
\affiliation{Department of Physics, University of Adelaide, Adelaide, 5005, Australia}
\affiliation{Dept. of Physics and Astronomy, University of Alaska Anchorage, 3211 Providence Dr., Anchorage, AK 99508, USA}
\affiliation{Dept. of Physics, University of Texas at Arlington, 502 Yates St., Science Hall Rm 108, Box 19059, Arlington, TX 76019, USA}
\affiliation{CTSPS, Clark-Atlanta University, Atlanta, GA 30314, USA}
\affiliation{School of Physics and Center for Relativistic Astrophysics, Georgia Institute of Technology, Atlanta, GA 30332, USA}
\affiliation{Dept. of Physics, Southern University, Baton Rouge, LA 70813, USA}
\affiliation{Dept. of Physics, University of California, Berkeley, CA 94720, USA}
\affiliation{Lawrence Berkeley National Laboratory, Berkeley, CA 94720, USA}
\affiliation{Institut f{\"u}r Physik, Humboldt-Universit{\"a}t zu Berlin, D-12489 Berlin, Germany}
\affiliation{Fakult{\"a}t f{\"u}r Physik {\&} Astronomie, Ruhr-Universit{\"a}t Bochum, D-44780 Bochum, Germany}
\affiliation{Universit{\'e} Libre de Bruxelles, Science Faculty CP230, B-1050 Brussels, Belgium}
\affiliation{Vrije Universiteit Brussel (VUB), Dienst ELEM, B-1050 Brussels, Belgium}
\affiliation{Department of Physics and Laboratory for Particle Physics and Cosmology, Harvard University, Cambridge, MA 02138, USA}
\affiliation{Dept. of Physics, Massachusetts Institute of Technology, Cambridge, MA 02139, USA}
\affiliation{Dept. of Physics and The International Center for Hadron Astrophysics, Chiba University, Chiba 263-8522, Japan}
\affiliation{Department of Physics, Loyola University Chicago, Chicago, IL 60660, USA}
\affiliation{Dept. of Physics and Astronomy, University of Canterbury, Private Bag 4800, Christchurch, New Zealand}
\affiliation{Dept. of Physics, University of Maryland, College Park, MD 20742, USA}
\affiliation{Dept. of Astronomy, Ohio State University, Columbus, OH 43210, USA}
\affiliation{Dept. of Physics and Center for Cosmology and Astro-Particle Physics, Ohio State University, Columbus, OH 43210, USA}
\affiliation{Niels Bohr Institute, University of Copenhagen, DK-2100 Copenhagen, Denmark}
\affiliation{Dept. of Physics, TU Dortmund University, D-44221 Dortmund, Germany}
\affiliation{Dept. of Physics and Astronomy, Michigan State University, East Lansing, MI 48824, USA}
\affiliation{Dept. of Physics, University of Alberta, Edmonton, Alberta, T6G 2E1, Canada}
\affiliation{Erlangen Centre for Astroparticle Physics, Friedrich-Alexander-Universit{\"a}t Erlangen-N{\"u}rnberg, D-91058 Erlangen, Germany}
\affiliation{Physik-department, Technische Universit{\"a}t M{\"u}nchen, D-85748 Garching, Germany}
\affiliation{D{\'e}partement de physique nucl{\'e}aire et corpusculaire, Universit{\'e} de Gen{\`e}ve, CH-1211 Gen{\`e}ve, Switzerland}
\affiliation{Dept. of Physics and Astronomy, University of Gent, B-9000 Gent, Belgium}
\affiliation{Dept. of Physics and Astronomy, University of California, Irvine, CA 92697, USA}
\affiliation{Karlsruhe Institute of Technology, Institute for Astroparticle Physics, D-76021 Karlsruhe, Germany}
\affiliation{Karlsruhe Institute of Technology, Institute of Experimental Particle Physics, D-76021 Karlsruhe, Germany}
\affiliation{Dept. of Physics, Engineering Physics, and Astronomy, Queen's University, Kingston, ON K7L 3N6, Canada}
\affiliation{Department of Physics {\&} Astronomy, University of Nevada, Las Vegas, NV 89154, USA}
\affiliation{Nevada Center for Astrophysics, University of Nevada, Las Vegas, NV 89154, USA}
\affiliation{Dept. of Physics and Astronomy, University of Kansas, Lawrence, KS 66045, USA}
\affiliation{Centre for Cosmology, Particle Physics and Phenomenology - CP3, Universit{\'e} catholique de Louvain, Louvain-la-Neuve, Belgium}
\affiliation{Department of Physics, Mercer University, Macon, GA 31207-0001, USA}
\affiliation{Dept. of Astronomy, University of Wisconsin{\textemdash}Madison, Madison, WI 53706, USA}
\affiliation{Dept. of Physics and Wisconsin IceCube Particle Astrophysics Center, University of Wisconsin{\textemdash}Madison, Madison, WI 53706, USA}
\affiliation{Institute of Physics, University of Mainz, Staudinger Weg 7, D-55099 Mainz, Germany}
\affiliation{Department of Physics, Marquette University, Milwaukee, WI 53201, USA}
\affiliation{Institut f{\"u}r Kernphysik, Westf{\"a}lische Wilhelms-Universit{\"a}t M{\"u}nster, D-48149 M{\"u}nster, Germany}
\affiliation{Bartol Research Institute and Dept. of Physics and Astronomy, University of Delaware, Newark, DE 19716, USA}
\affiliation{Dept. of Physics, Yale University, New Haven, CT 06520, USA}
\affiliation{Columbia Astrophysics and Nevis Laboratories, Columbia University, New York, NY 10027, USA}
\affiliation{Dept. of Physics, University of Oxford, Parks Road, Oxford OX1 3PU, United Kingdom}
\affiliation{Dipartimento di Fisica e Astronomia Galileo Galilei, Universit{\`a} Degli Studi di Padova, I-35122 Padova PD, Italy}
\affiliation{Dept. of Physics, Drexel University, 3141 Chestnut Street, Philadelphia, PA 19104, USA}
\affiliation{Physics Department, South Dakota School of Mines and Technology, Rapid City, SD 57701, USA}
\affiliation{Dept. of Physics, University of Wisconsin, River Falls, WI 54022, USA}
\affiliation{Dept. of Physics and Astronomy, University of Rochester, Rochester, NY 14627, USA}
\affiliation{Department of Physics and Astronomy, University of Utah, Salt Lake City, UT 84112, USA}
\affiliation{Dept. of Physics, Chung-Ang University, Seoul 06974, Republic of Korea}
\affiliation{Oskar Klein Centre and Dept. of Physics, Stockholm University, SE-10691 Stockholm, Sweden}
\affiliation{Dept. of Physics and Astronomy, Stony Brook University, Stony Brook, NY 11794-3800, USA}
\affiliation{Dept. of Physics, Sungkyunkwan University, Suwon 16419, Republic of Korea}
\affiliation{Institute of Basic Science, Sungkyunkwan University, Suwon 16419, Republic of Korea}
\affiliation{Institute of Physics, Academia Sinica, Taipei, 11529, Taiwan}
\affiliation{Dept. of Physics and Astronomy, University of Alabama, Tuscaloosa, AL 35487, USA}
\affiliation{Dept. of Astronomy and Astrophysics, Pennsylvania State University, University Park, PA 16802, USA}
\affiliation{Dept. of Physics, Pennsylvania State University, University Park, PA 16802, USA}
\affiliation{Dept. of Physics and Astronomy, Uppsala University, Box 516, SE-75120 Uppsala, Sweden}
\affiliation{Dept. of Physics, University of Wuppertal, D-42119 Wuppertal, Germany}
\affiliation{Deutsches Elektronen-Synchrotron DESY, Platanenallee 6, D-15738 Zeuthen, Germany}

\author[0000-0001-6141-4205]{R. Abbasi}
\affiliation{Department of Physics, Loyola University Chicago, Chicago, IL 60660, USA}

\author[0000-0001-8952-588X]{M. Ackermann}
\affiliation{Deutsches Elektronen-Synchrotron DESY, Platanenallee 6, D-15738 Zeuthen, Germany}

\author{J. Adams}
\affiliation{Dept. of Physics and Astronomy, University of Canterbury, Private Bag 4800, Christchurch, New Zealand}

\author[0000-0002-9714-8866]{S. K. Agarwalla}
\altaffiliation{also at Institute of Physics, Sachivalaya Marg, Sainik School Post, Bhubaneswar 751005, India}
\affiliation{Dept. of Physics and Wisconsin IceCube Particle Astrophysics Center, University of Wisconsin{\textemdash}Madison, Madison, WI 53706, USA}

\author[0000-0003-2252-9514]{J. A. Aguilar}
\affiliation{Universit{\'e} Libre de Bruxelles, Science Faculty CP230, B-1050 Brussels, Belgium}

\author[0000-0003-0709-5631]{M. Ahlers}
\affiliation{Niels Bohr Institute, University of Copenhagen, DK-2100 Copenhagen, Denmark}

\author[0000-0002-9534-9189]{J.M. Alameddine}
\affiliation{Dept. of Physics, TU Dortmund University, D-44221 Dortmund, Germany}

\author{N. M. Amin}
\affiliation{Bartol Research Institute and Dept. of Physics and Astronomy, University of Delaware, Newark, DE 19716, USA}

\author[0000-0001-9394-0007]{K. Andeen}
\affiliation{Department of Physics, Marquette University, Milwaukee, WI 53201, USA}

\author[0000-0003-4186-4182]{C. Arg{\"u}elles}
\affiliation{Department of Physics and Laboratory for Particle Physics and Cosmology, Harvard University, Cambridge, MA 02138, USA}

% ...

\begin{abstract}


Active Galactic Nuclei (AGN) are prime candidate sources of the high-energy, astrophysical neutrinos detected by IceCube. 
This is demonstrated by the real-time multi-messenger detection of the blazar TXS 0506+056  and the recent evidence of neutrino emission from NGC 1068 from a separate time-averaged study. 
However, the production mechanism of the astrophysical neutrinos in AGN is not well established which can be resolved via correlation studies with photon observations. 
For neutrinos produced due to photohadronic interactions in AGN, in addition to a correlation of neutrinos with high-energy photons, there would also be a correlation of neutrinos with photons emitted at radio wavelengths.
In this work, we perform an in-depth stacking study of the correlation between 15 GHz radio observations of AGN reported in the MOJAVE XV catalog, and ten years of neutrino data from IceCube.  We also use a time-dependent approach which improves the statistical power of the stacking analysis.
No significant correlation was found for both analyses and upper limits are reported. When compared to the IceCube diffuse flux, at 100 TeV and for a spectral index of 2.5, the upper limits derived are $\sim3\%$ and $\sim9\%$ for the time-averaged and time-dependent case, respectively.


\end{abstract}

%
%

\section{Introduction}
\label{sec:intro}

Neutrinos are a valuable complementary messenger to photons, however, their elusive nature adds complexities to their detection leading to uncertainties about the exact sources producing them. 
Understanding the neutrino production mechanism and the concurrent detection of photons can help pinpoint their sources. This can be done using a hypothesis that the observed neutrinos and photons follow a certain correlation. This correlation supports the theory that both particles originate from similar or related processes within or around extragalactic sources. 
A positive correlation will help us find the sources that may be producing neutrinos and understand the processes that lead to their creation. 
However, these correlation studies are limited by a lack of coincident photon observations with neutrino data reducing their sensitivities. 
One of the theories involving neutrino production in Active Galactic Nuclei (AGN) involves the possibility of a correlation with photon detected at radio wavelengths (see Sec~\ref{sec:neutrino_production_in_AGN}).
In this work, we perform a stacking analysis using 10 years of data collected by the IceCube Neutrino Observatory (2008-2018) along with the time-dependent photon observations published in the MOJAVE XV catalog \citep{MOJAVEXV} to test for correlation between the radio and neutrino observations. The time-dependent stacking study makes use of additional coincident information and improves the statistical power of traditional stacking analyses.

The IceCube Neutrino Observatory, located at the geographic South Pole, is a cubic kilometer in-ice neutrino detector that has collected $\sim$18 years of neutrino data \citep{icecube_2017_intro}. When high-energy neutrinos pass through the Earth, they may interact with the ice or surrounding bedrock, creating secondary charged particles. 
These particles produce Cherenkov light which is detected and used to reconstruct the high-energy neutrino interaction energy and direction. 
While the first evidence of astrophysical neutrino diffuse flux detection was reported in 2013 \citep{icecube_2013_astrophysical_neutrino}, the origin of these elusive particles and the sources producing them remains uncertain. One of the prime candidates for the origin of these particles are AGN, which are active supermassive black holes, some of which have jets of extremely high-energy particles originating from the center \citep{AGN_potential_neutrino_1979,AGN_potential_neutrino_1981}. 

In 2017, IceCube detected a high-energy neutrino event in a direction coincident with the AGN TXS 0506+056\footnote{https://gcn.gsfc.nasa.gov/gcn3/21916.gcn3} which was found to be flaring in gamma rays \citep{TXS_Icecube_fermi_others_multimessenger}. A follow-up analysis of archival IceCube neutrino data revealed an earlier burst of neutrino events from the same source in 2014/2015 without an accompanying flare of gamma rays \citep{TXS_Icecube}. This source is a blazar which is a type of AGN with the jet pointed in the direction of the observer. Recently, \cite{ngc1068_22} reported significant evidence of neutrino emission from NGC 1068, a nearby Seyfert II type of AGN. 
Seyfert II sources are galaxies that are observed with narrow emission lines in their spectrum and a variable radio emission. 
Some Seyfert II AGN also have jets radiating outward, however, they are relatively dim gamma-ray sources, unlike blazars which have relativistic jets oriented in Earth’s direction.
These detections motivated studies involving all AGNs as potential neutrino sources. To better explain neutrino production in these sources, theoretical modeling efforts and correlation studies are performed with photon observations in a particular energy regime.
Multiple independent analyses from both: the IceCube Collaboration \citep[see for example][]{3fhl_icecube,icrc_mojave, federica_agn_cores_22,christina_alert_plavin_fermi} and other researchers \citep[see for example][]{Plavin_2020a,  Plavin_2021,Kun_radio_neutrino_external_2022,Rodrigues_2021,  zhou2021neutrino_agn} which used different datasets and analysis methodology were performed. 
Based on these studies and AGN model predictions, one of the leading theories is that the neutrinos observed by IceCube may be correlated with the photons observed in the radio regime as the radio variability traces environment conditions ideal for neutrino production \citep[See][and Sec~\ref{sec:neutrino_production_in_AGN} of this work for more details] {agn_pp_pgamma_Jacobsen,murase_stecker_agn_review}. 
Additionally, as blazar jets point in the direction of the observer, due to Doppler boosting the broadband SED of radio-loud AGN is not clear with the jet emission dominating over the electromagnetic signal. As the neutrino signal from these  AGNs will not be affected by the jets, a positive correlation will help identification of radio-loud AGN accelerators and understand their relevant neutrino production processes. 


\subsection{Neutrino-radio correlation in AGN}
\label{sec:neutrino_production_in_AGN}

Theoretical predictions of neutrino production in AGNs depend on the the type of interaction and the location where it can occur. Two types of interaction that can lead to the production of these neutrinos from AGN are the photohadronic (nucleon-photon or $p\gamma$) and hadronuclear (nucleon-nucleon or $pp$) processes, which can occur close to the core of the AGN or in the jet of a jetted-AGN \citep{pp_suppression_1987_sikora,agn_pgamma_Stecker,murase_stecker_agn_review}. 
Depending on where these processes take place, there may be a correlation of the neutrino signal with photon observations at a particular wavelength. Understanding this correlation, or a lack of it, can help researchers pinpoint the neutrino production mechanism.

Here we focus on testing the theory behind the neutrino and photon correlations in AGN. 
The synchrotron radiation resulting from accelerated electrons leads to the emission of photons, observable at radio frequencies, which in turn undergo Inverse Compton scattering to form the X-ray photons. In the case of neutrino production due to photohadronic interactions, these X-ray photons interact with protons to give pions that decay to give gamma rays and neutrinos. For opaque or obscured AGN, the resultant gamma-ray photons will cascade down to lower energies (e.g. X-rays) before escaping from the core of the AGN \citep{Murase_2022_hidden_heart_AGN}. Neutrinos, on the other hand, escape without interaction for both obscured and un-obscured AGN.
While this suggests a possible correlation of neutrino observations with photons detected in the X-ray regime for certain AGN, it also supports a correlation of neutrinos with photons observed in the radio regime for all AGN.
This is because, if an AGN is flaring at radio wavelengths, it can signify an increase in the conditions favorable for the neutrino production process. This means that the neutrino signal will be directly correlated with the radio flux density measurements of the AGN \citep[see][for more details]{agn_pp_pgamma_Jacobsen}. Additionally, as compared to X-ray observations, radio detection of photons from AGN is easier and has been carried out by multiple radio observatories over time, allowing researchers to use archival data of photon observations. Studies like \cite{Plavin_2020a} report that radio measurements at higher frequencies ($>10\,$GHz) are correlated with neutrino events showing increased emission activity as compared to lower frequencies. To test this correlation, in this work, we use the MOJAVE XV dataset \citep{MOJAVEXV} which reports 15\,GHz observations of AGN sources observed over twenty years.

\subsection{Previous AGN correlation studies}
\label{sec:AGN_correlation_studies}
Various observational and theoretical papers 
exist to study photon and neutrino correlations, however, it is unclear what are the dominant processes for neutrino production in AGN  \citep[see review by][]{murase_stecker_agn_review}. 
While studies like \cite{Plavin_2021},\cite{OVRO_neutrinos_2020} and \cite{Buson_2022} report 
a correlation of IceCube neutrinos with photons from AGN, other studies like  \cite{zhou2021neutrino_agn},\cite{federica_agn_cores_22} and \cite{christina_alert_plavin_fermi}, do not find a significant correlation. 
One of the factors that contribute to the discrepancy is the lack of data, which, in terms of neutrinos can be the usage of the IceCube alert dataset instead of the full IceCube tracks dataset \citep{icecube_10yr_data}.
As an example, the all-sky point source neutrino dataset made of track events \citep{icecube_10yr_data} has more muon track-like events (see Sec~\ref{sec:icecube_data}) as a function of energy along with a better, more consistent coverage in time and location as compared to the IceCube public alert sample \citep{IceCube_alert_icecat1_23}. For a source with an assumed power-law spectrum with an index of $\gamma = 2.5$, the public alert selection has a factor of $\times$100 fewer astrophysical neutrinos than the full selection. 
Another factor that can affect the discrepancy between the neutrino studies can be 
the inclusion of additional components like energy information in signal PDFs or signalness for neutrino events \citep[see for example][]{federica_agn_cores_22,christina_alert_plavin_fermi}.
Finally, apart from flaring AGN analyses, correlation studies involving stacking are often limited to using time-averaged measurements. This is due to the lack of time-dependent photon and neutrino data. Including time-dependent information in a stacking study will allow us to weight sources based on the photon data as a function of time and increase the sensitivity of the analysis.  
It thus becomes important to use the most complete IceCube neutrino dataset in combination with a large, reasonably complete AGN source catalog preferably including time-dependent flux information. 




\subsection{Paper Outline}
\label{sec:paper_outline}
This work uses the 10-year IceCube tracks dataset using a stacking analysis \citep[similar to ][]{Braun_2008,icrc_mojave,federica_agn_cores_22} to search for correlation between radio and neutrino data from AGN. The stacking analysis is performed for both the time-averaged and time-dependent cases assuming there is a 1:1 correlation between the 15\,GHz radio flux density and the high-energy neutrino flux.
The article is divided as follows: Section.~\ref{sec:data_samples} describes in detail the neutrino and photon datasets used in this study; Section.~\ref{sec:analysis_method} describes the likelihood framework used;  Section.~\ref{sec:results} presents the results derived from this work; and Section.~\ref{sec:conclusion} discusses the conclusion and future implications of this work.

%
%

\begin{figure*}
\centering
\includegraphics[width=.7\textwidth,trim={0.0cm 2.8cm 0.0cm 0.0cm}]{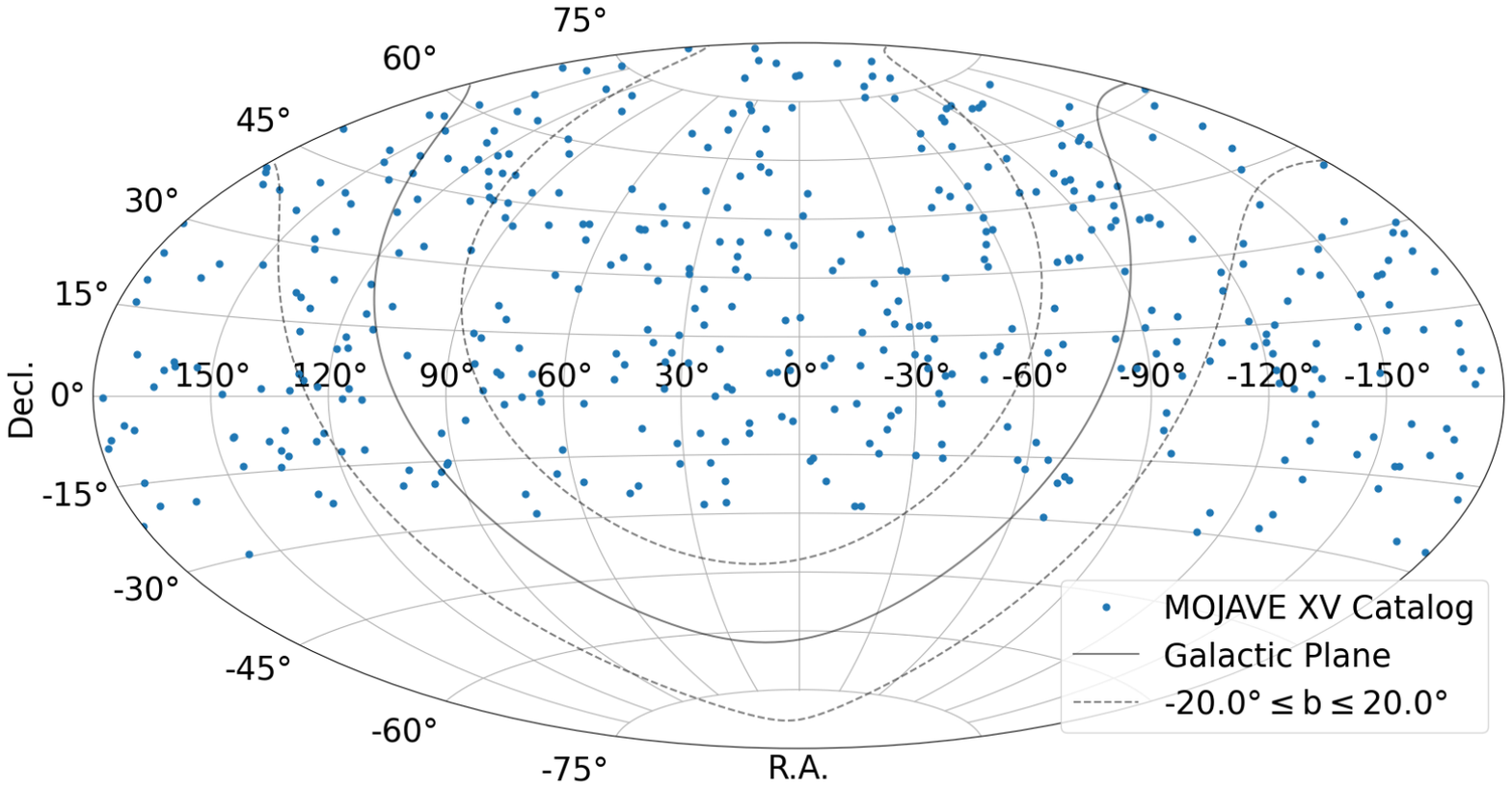}
\caption{Source distribution of MOJAVE XV catalog using Equatorial (J2000) coordinates. Note that the source distribution of the MOJAVE catalog covers the equatorial and northern hemispheres where IceCube’s sensitivity for track-like events is maximized \citep[see sensitivity curve reported in ][]{10yrPStracks_tessa}.
}
\label{fig:aitoff}
\end{figure*}

\section{Data Samples}
\label{sec:data_samples}

\subsection{IceCube Selection of Track-Like Events}
\label{sec:icecube_data}

IceCube records neutrino events following a signature of being \enquote{track-like} or \enquote{cascade-like} depending on the type of particle interaction leading to the detection. 
Charged-current muon neutrino interactions lead to elongated  \enquote{track-like} signatures produced due to long-lived muons that travel several kilometers through the ice. On the other hand, neutral-current interactions or charged current electron and tau neutrino interactions produce hadronic and electromagnetic showers covering a smaller distance \citep{cascade_topology_halzen} giving \enquote{cascade-like} signatures. This work focuses on \enquote{track-like} events as they have a better directional resolution as compared to \enquote{cascade-like} events.
The neutrino data is obtained using the all-sky point-source tracks sample \citep{icecube_10yr_data}, which spans over a duration of 10 years, from April 2008 to July 2018. 
The properties of this neutrino point-source events sample along with the selection and filtering methods are described in detail by \cite{10yrPStracks_tessa}. 

The ten-year dataset is further divided by the configuration of IceCube detector strings. The first year (2008) dubbed \enquote{IC-40} uses data from the 40 string configuration, \enquote{IC-59} (2009) uses the 59 string configuration, \enquote{IC-79} (2010) uses the 79 string configuration and \enquote{IC-86} (2011-2018) uses the 86 string configuration. For more details regarding the configurations see \cite{icecube_10yr_data}.



\subsection{MOJAVE Data of AGN at 15\,GHz}
\label{sec:mojave_data}

The AGN sources used for this study are taken from the MOJAVE XV catalog which includes the total flux density observations of 437 sources obtained with the Very Long Baseline Array (VLBA) at 15\,GHz \citep{MOJAVEXV}. The catalog consists of a total of 5321 observations of these AGN made with varying cadence and number of observations per AGN obtained between 1996 January 19 and 2016 December 26. The MOJAVE XV AGN catalog is a blazar-dominated sample with 392 blazar sources, 27 radio galaxies, 13 unidentified AGN and 5 narrow-line Seyfert I galaxies. The MOJAVE source list was updated over time to include low redshift radio galaxies with spectra peaking in the GHz regime. These sources are distributed almost uniformly over the sky at declinations $\delta > -30\arcdeg$ ((see Fig.~\ref{fig:aitoff}) which is better matched to the improved sensitivity of the work presented here in the northern hemisphere \citep[see sensitivity curve reported in ][]{10yrPStracks_tessa}.  Moreover, all the observed AGN have bright compact radio emissions with total flux densities greater than 50\,mJy. This implies that the observed sources are bright at 15\,GHz and changes of the radio emission can be effectively measured. This makes the MOJAVE catalog one of the most promising radio catalogs for correlation studies such as this one. 

\begin{figure*}
\centering
\includegraphics[width=.6\textwidth,trim={2.0cm 1.0cm 2.0cm 1.0cm}]{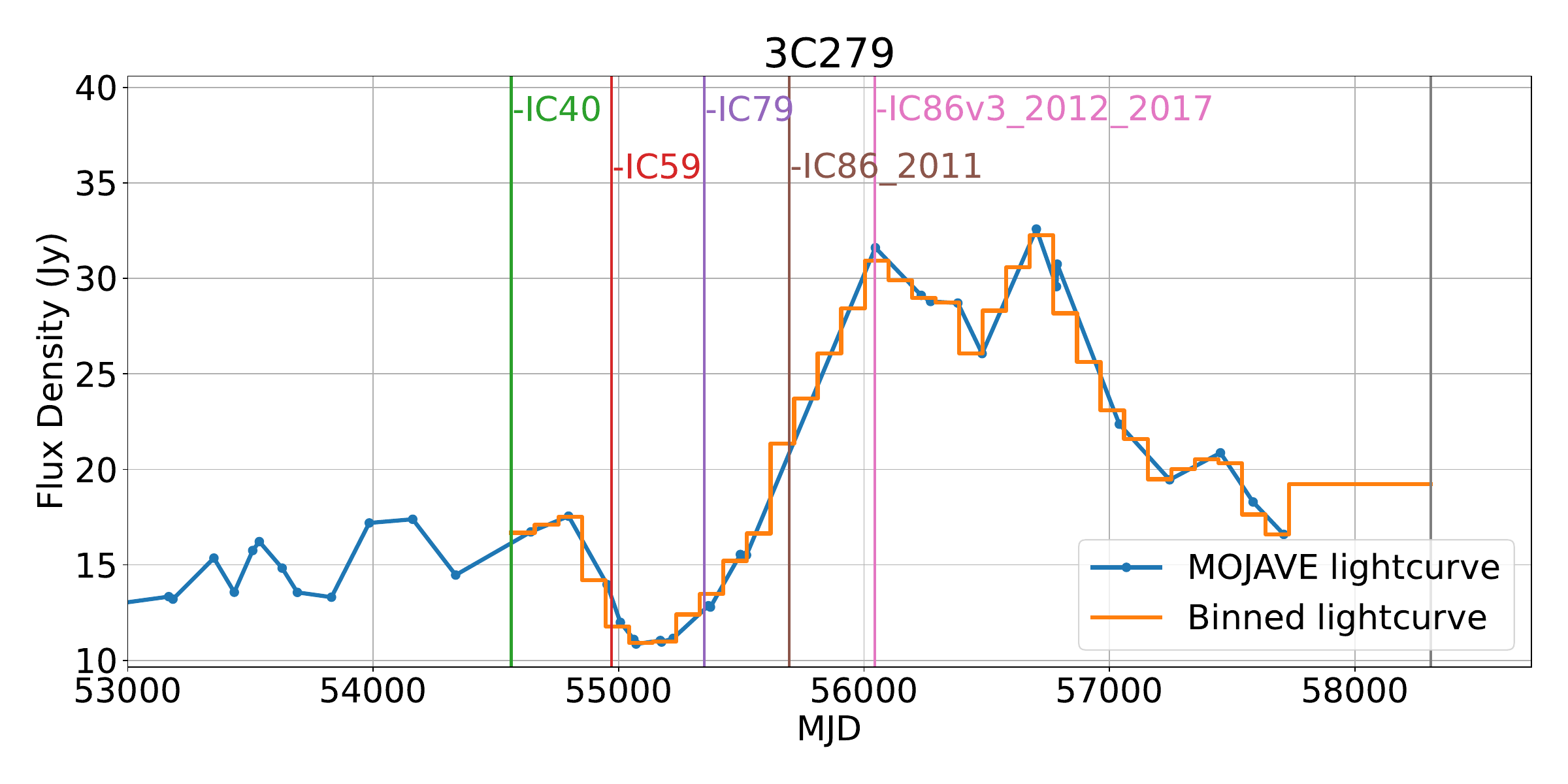}
\caption{
Example binned lightcurve (orange line) using MOJAVE data (blue data points) for one source (3C279) along with different IceCube datasets, and their time coverage is shown with the corresponding dataset name. The light curve is binned using equal time binning within the 10-year IceCube data shown here by the vertical lines. 
Left to right, the bins are shown as:
green to red (2008-2009;IC40), 
red to purple (2009-2010;IC59),
purple to brown (2010-2011;IC79),
brown to pink (2011-2012;IC86\_2011),
pink to grey (2012-2017;IC86v3\_2012\_2017).
}

\label{fig:lc_example}
\end{figure*}

As this study also uses multi-epoch observations in the form of photon flux density light curves, the MOJAVE XV catalog is preferred over the Radio Fundamental Catalog (RFC)\footnote{http://astrogeo.org/rfc/}. The RFC catalog is more complete but lacks time-dependent radio measurements. While many sources in the MOJAVE XV catalog are AGN that are consistently observed with a good cadence (see, e.g. Fig.~\ref{fig:lc_example}), there do exist some sources that have not been observed frequently. We remove the sources that have very few observations (keeping sources with a minimum of three observations), reducing the sample size of 437 sources to 397 sources.  

The MOJAVE sample is considered to be complete in terms of sources observed at 15\,GHz by VLBA and with flux densities with $f_\nu > 1.5$\,Jy. However, a completeness correction is required for larger, unbiased analyses such as this, to account for the sources not included either due to spatial coverage or flux threshold in the catalog. The completeness correction is found to be $44.7\pm11.2\%$ (see Appendix A).
After accounting for completeness, this study focuses on a blazar-dominated AGN sample that follows the same properties as the sources in the MOJAVE sample.

%
%
\section{Analysis Method}
\label{sec:analysis_method}

Using the neutrino and radio flux density described above, we search for cross-correlations between neutrinos and photon observations in the direction of the AGN in the MOJAVE XV catalog. This is done by using a stacking analysis weighted according to the observed time-averaged and time-dependent radio flux density. The basis of this work is a likelihood approach similar to the ones described by \cite{Braun_2008} 
and others, where the track-like neutrino data is modeled using a background hypothesis ($B_i$) and signal hypothesis ($S_i$). The background data consists of atmospheric events from neutrinos and muons originating in the Earth's atmosphere, while the signal consists of a radio source-associated point-like excess of neutrinos in the stacked data. 

As described in Sec.~\ref{sec:icecube_data}, the point source tracks sample is further divided into five configurations depending on the number of strings in use along with selections, software and calibrations used \citep[see Table I. of][]{icecube_10yr_data}. 
If each configuration is denoted by an index $k$, let $N_k$ be the number of total neutrino events in the configuration. Using the notation of $n_s$ for the number of signal events from a certain direction and $N_k$ for the number of events in the configuration along with the signal and background hypothesis, a likelihood function can be constructed for each event $i$ by the following: 
\begin{equation}
    \mathcal{L}(n_s, \gamma)=\prod_k\prod_i^{N_k}\left(\frac{n_s^k}{N_k}{S}_i^k  +\left(1-\frac{n_s^k}{N_k}\right){B}_i^k \right)
    \label{eq:likelihood}
\end{equation}
where ${S}_i^k$ and ${B}_i^k$ represent the probability density functions (PDFs) corresponding to the signal and background hypotheses, respectively, for each configuration $k$. The expected number of signal events $n_s^k$ is derived by computing the fraction of the total events $n_s$ in a configuration $k$ using a factor $f_k$ where $n_s^k=f_k n_s$. The fractional contribution $f_k$ and ${S}_i^k$ are modified for a stacking analysis to include per-source weighting information. If, for a configuration $k$, $\omega_j$ is the per-source contribution for the model being tested and ${R}_j^k$ is the detector weight at the declination of the $j^{th}$ source, $f_k$ and ${S}_i^k$ can be written as:
\begin{equation}
    {S}_i^{k}= \frac{\sum_j \omega_j {R}_j^k {S}^k_{ij}}{\sum_j \omega_j{R}_j^k}
    \label{eq:sig_weighting}
\end{equation}
and
\begin{equation}
    f_k= \frac{\sum_j \omega_j {R}_j^k}{\sum_k\sum_j \omega_j{R}_j^k}
    \label{eq:frac_weighting}
\end{equation}
The detector weights ${R}_j^k$ vary per source and account for the detection efficiency of IceCube. The detection efficiency for signal events depends on the source direction from which the signal neutrinos originate along with the differential spectrum (assumed as $E^{-\gamma}$) of the neutrinos and is calculated using Monte Carlo simulations \citep{mhuber_thesis_blazar_stacking}. The source weights $\omega_j$ depend on the hypothesis being tested.
In this work, these parameters are determined according to the special cases of time-averaged stacking and time-dependent stacking described below.

\subsection{Time-averaged Stacking}
\label{sec:time_integrated}

The average flux densities of each source in the MOJAVE XV catalog are used as weights $\omega_j$ for the stacking in Eqs.~\ref{eq:sig_weighting} and \ref{eq:frac_weighting} under the assumption that there is a 1:1 correlation between the radio flux density and IceCube neutrino flux. For an astrophysical source at direction $\vec{x_j}$, using a set of neutrino data events, indexed by $i$, each with reconstructed energy and direction given by $E_i$ and $\vec{x_i}$ respectively, we create a power-law energy distribution $P(E_i|\gamma)$ where $\gamma$ indicates the spectral index \citep[see also][]{Braun_2008}. 

The signal hypothesis for a source, denoted by index j, is modeled using 
\begin{equation}
    {P}^k_{sig}(\sigma_i,x_i,x_j) =  \frac{1}{2\pi\sigma_i^2}\exp\left({-\frac{|\vec{x_i}-\vec{x_j}|^2}{2\sigma_i^2}}\right) 
\end{equation}
where $\sigma_i$ is the angular reconstruction error estimate. Combining this information with the energy-dependent signal PDF ($\epsilon_i^{k}$) we get the $S_{ij}^k$ term, given as:
\begin{equation}
    {S}_{ij}^k= {P}^k_{sig}(\sigma_i,x_i,x_j) \,\,  \epsilon_i^{k}(E_i,\delta_i|\gamma) 
\end{equation}
where $\epsilon_i^{k}$ is computed by using a power-law energy spectrum with index $\gamma$. This is then used in Eq.~\ref{eq:sig_weighting} to account for the per-source weighting for the stacking.
The directional uncertainty for the event reconstruction in the configurations uses a lower limit of $0.2\arcdeg$ to minimize the impact of any inaccuracies in ice models and to ensure the likelihood calculation is not dominated by a single event \citep{icecube_10yr_data}. Note that the radio source position uncertainty (in the order of milli-arcseconds) is negligible compared to the above lower limit on the directional uncertainty of the neutrino event reconstructions.

Finally, the background hypothesis is modeled using:
\begin{equation}
{B}_i^k(\vec{x_i},E_i,\delta_i) =  {P}_{bkg}^k(\vec{x_i}) \epsilon_B^k(E_i,\delta_i)
\end{equation}
where the product of the energy-dependent background PDF $\epsilon_B^k$ is taken with the spatial PDF ${P}_{bkg}^k$ at declination $\delta_i$. No per-source weighting is applied to the background hypothesis.

\subsection{Time-dependent Stacking}
\label{sec:time_dependent}

The time-dependent stacking analysis differs slightly from the time-averaged case, where we test a 1:1 correlation between the time-dependent radio flux density measurements at 15\,GHz with the neutrino flux seen by IceCube. For this case, separate light curves are created for each source using the flux density measurements. These light curves are then binned over the 10-year neutrino data period. The number of bins is kept fixed at 40 bins for each source. Based on tests with multiple bin values, the 40 bin value is chosen to ensure no temporal information is lost per AGN and the width and location of the bin height are the same for all sources while constraining the computational limit required to perform the analysis. Increasing the number of bins does not give any additional increase in sensitivity but significantly increases the computing power required while decreasing the bins leads to reduced sensitivity due to a loss of lightcurve information in some sources. An example of a binned light curve is shown in Fig.~\ref{fig:lc_example}. The weighting term for Eqs.~\ref{eq:sig_weighting} and ~\ref{eq:frac_weighting} in this case is given by the flux density of the $j^{th}$  source at time $t_i$. If a source has no observation during the observed period, the time-averaged flux density measurement is used (see Fig.~\ref{fig:lc_example}). This correction was generally applied for either very small time periods of the light curve or for less variable sources with fewer data points. In both cases, this does not impact the analysis significantly. 

Next, the signal and background hypothesis \citep[described above and in ][]{Braun_2008} is modified to include the temporal information using different time bins. For every time bin, the signal and background PDFs are computed, changing the equations to model the hypotheses to:
\begin{equation}
    {S}_{ij}^k(t)= {P}^k_{sig}(\sigma_i,x_i,x_j) \epsilon_i^{k}(E_i,\delta_i|\gamma) T^k_{sig.j}(t_i)
\end{equation}
and 
\begin{equation}
{B}_i^k(\vec{x_i},E_i,\delta_i,t) =  {P}_{bkg}^k(\vec{x_i}) \epsilon_B^k(E_i,\delta_i) T^k_{bkg}(t_i)
\end{equation}
where $T^k_{sig.j}$ and $T^k_{bkg}$ gives the temporal PDF at a time $t_i$ corresponding to bin $i$. Past analyses like \cite{tdep_icecube_qinrui} use a likelihood approach to search for flares by fitting time-dependent delay and threshold parameters which tests the possibility of a correlation with temporal delays or signal thresholds in a non-stacking approach. This study makes use of time-dependent radio flux densities for stacking testing a 1:1 correlation while not including the time-dependent delay and threshold parameters and fixing them to 0.   

Based on the signal and background hypothesis for the two stacking cases, along with the fractional contribution, the likelihood is calculated using Eq.~\ref{eq:likelihood}.
The test statistic (TS), which quantifies the significance of the results, is computed by making use of the likelihood formation in the following manner:
\begin{equation}
   TS = -2 \, {\rm sign}(\hat{n}_s) \log\left[ \frac{\mathcal{L}(\vec{x_s},n_s=0)}{\mathcal{L}(\vec{x_s},\hat{n}_s,\hat{\gamma})} \right] \, .
\end{equation} 
where the $\hat{}$ notation is used to denote a best-fit and $\vec{x_s}$ shows the stacked source position.

%
%
\section{Results and Discussion}
\label{sec:results}

\begin{figure}
\centering
  \centering
  \includegraphics[width=1.0\linewidth,trim={0.0cm 2.5cm 0.0cm 1.0cm}]{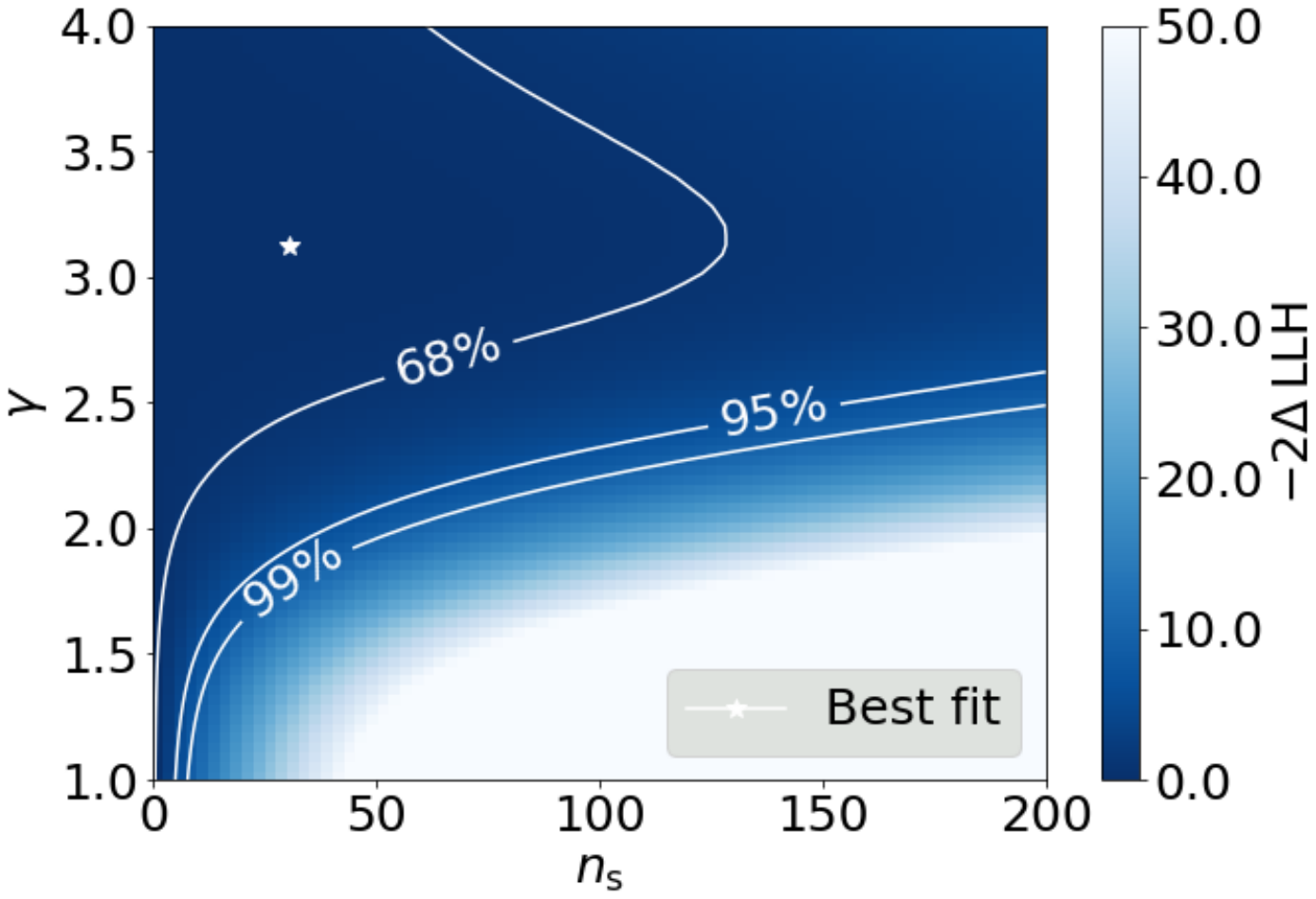}
  \centering
  \includegraphics[width=1.0\linewidth,trim={0.0cm 2.5cm 0.0cm 1.0cm}]{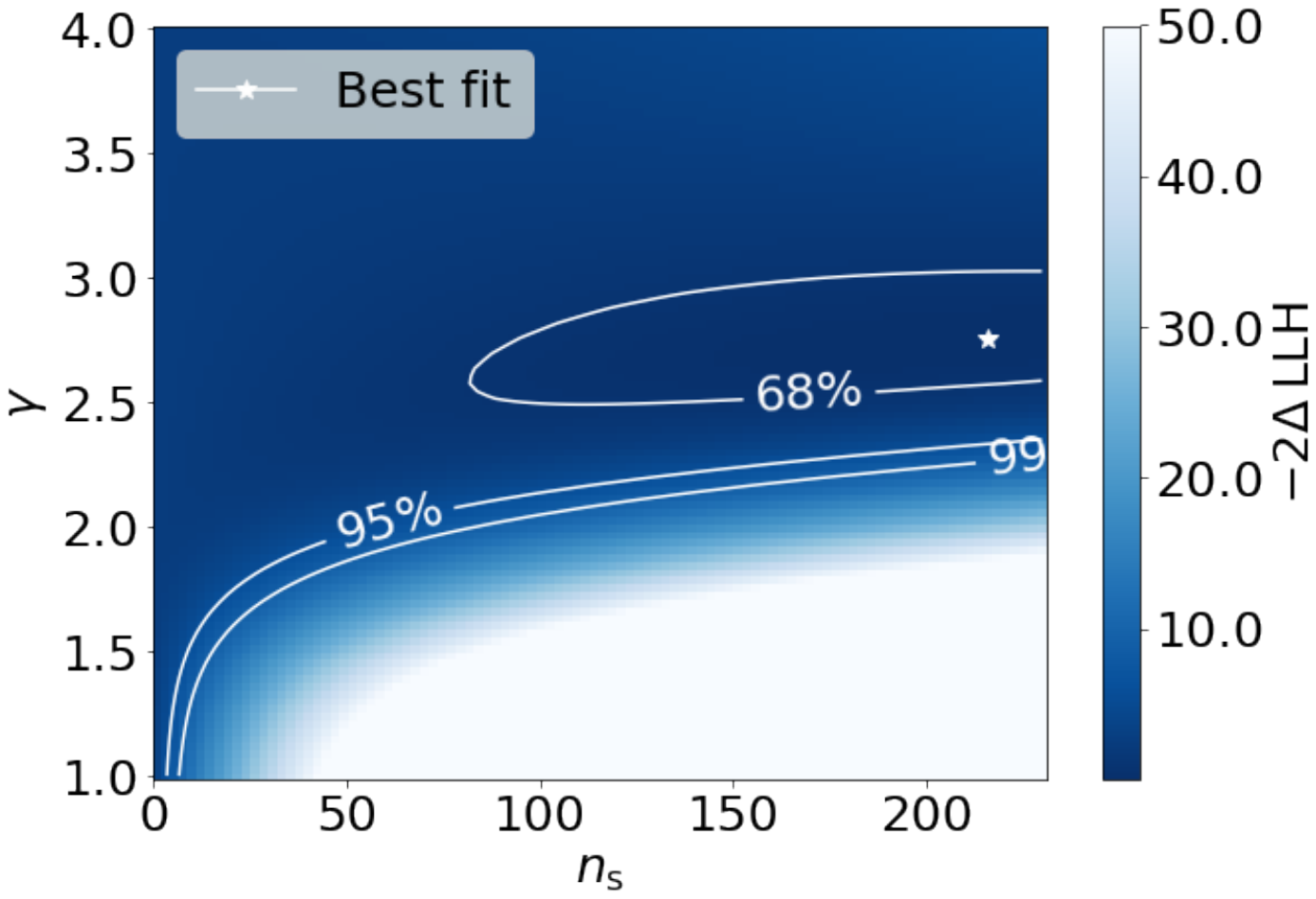}
\caption{Likelihood maps for the $n_s$-gamma values for the two analyses are shown here: time-averaged (top), time-dependent (bottom). The plot also shows the 1, 2 and 3 sigma confidence intervals from the best fit, assuming Wilk's theorem with 2 degrees of freedom.}
\label{fig:2dlikelihood}
\end{figure}


\subsection{Stacking results:}

For the time-averaged analysis, we obtain a best-fit spectral index of 3.14 and a best-fit number of signal events of 31.5. These best-fit results correspond to a p-value of 0.49  ($0.03\sigma$ significance) and the likelihood scan is shown in Fig.~\ref{fig:2dlikelihood} (top). 
Similarly, for the time-dependent analysis, we obtain a best-fit spectral index of 2.7 and a best-fit number of signal events of 218.7, corresponding to a p-value of 0.07 ($1.51\sigma$ significance). The likelihood contours for these best fits are shown in Fig.~\ref{fig:2dlikelihood}.  
To see how the best-fit TS scales with respect to the background distribution, see Fig~\ref{fig:background_dist} in the Appendix.
We can see that the best-fit number of signal events $\hat{n_s}$ derived for the time-dependent case is much higher than the time-averaged scenario with the time-averaged scenario having a softer spectral index value. While a 1:1 comparison for the parameters is difficult because of the change in the best fit spectral index, we can note the increase in the TS value of the fit for the time-dependent case. One can also see the changes in the shape of the contours shown in  Fig.~\ref{fig:2dlikelihood} with the time-dependent analysis appearing to constrain the fit in a better manner.
We report the study's upper limits, derived using the best-fit values, as we do not obtain a statistically significant result for both analyses. 

\begin{deluxetable}{lcc}
    \tablewidth{0pt}
    \tablecaption{Best-fit results derived from the study}
    \label{table:result}
    \tablehead{\colhead{} & Time-Averaged  & Time-Dependent }
    \startdata 
    Signal Events $\hat{n_s}$ & 31.5 & 218.7 \\
    Spectral Index $\hat{\gamma}$ & 3.1 & 2.7 \\
    TS & 0.09 & 2.58 \\
    p-value & 0.49 & 0.07 \\
    Significance & 0.03$\sigma$ & 1.51$\sigma $
    \enddata
\end{deluxetable}

\begin{figure*}[ht!]
   \begin{center}
   \begin{tabular}{c}
    \includegraphics[width=.49\textwidth,trim={0.0cm 1.8cm 0.0cm 0.0cm}]{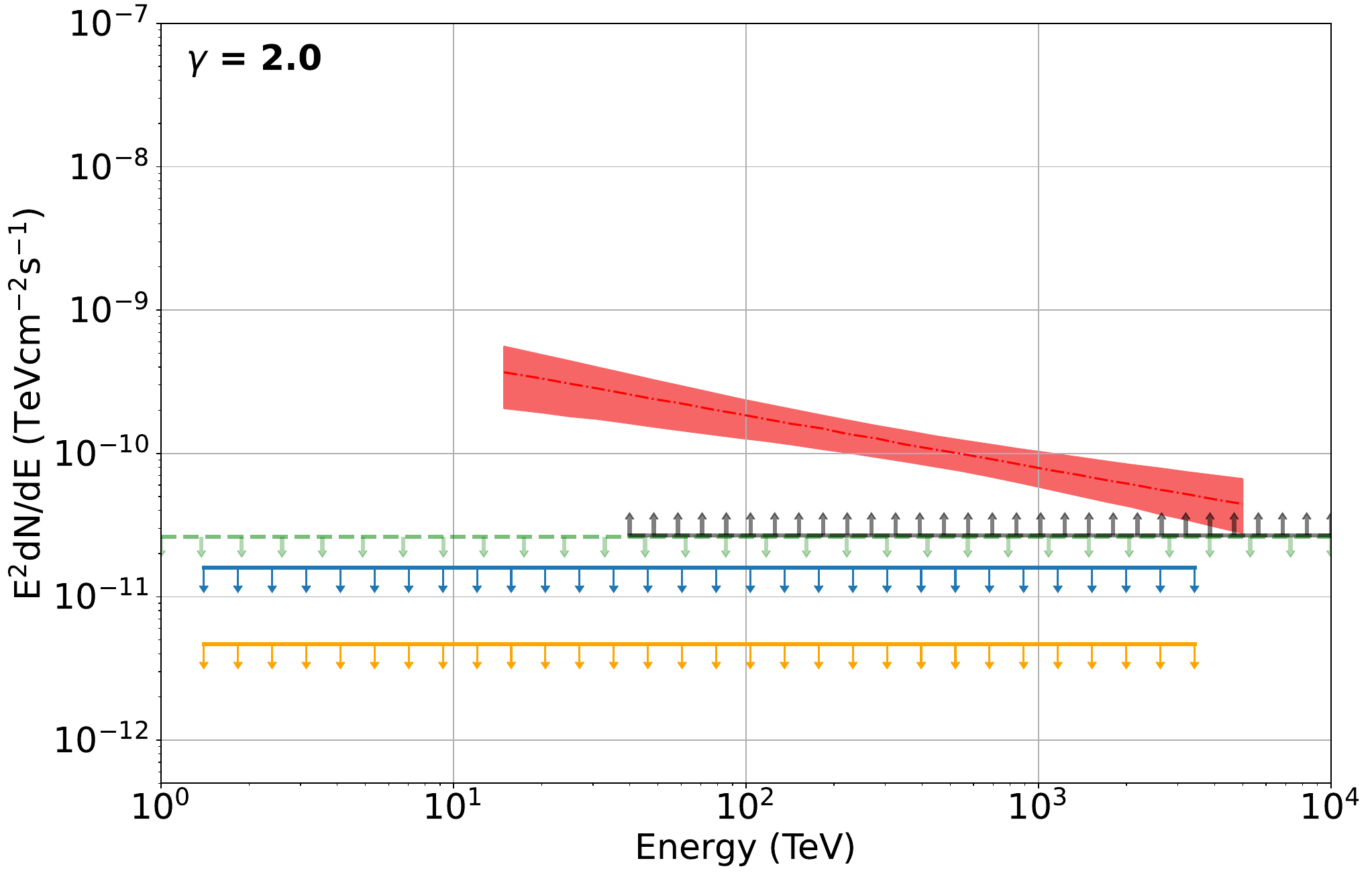}
    \includegraphics[width=.49\textwidth,trim={0.0cm 1.8cm 0.0cm 0.0cm}]{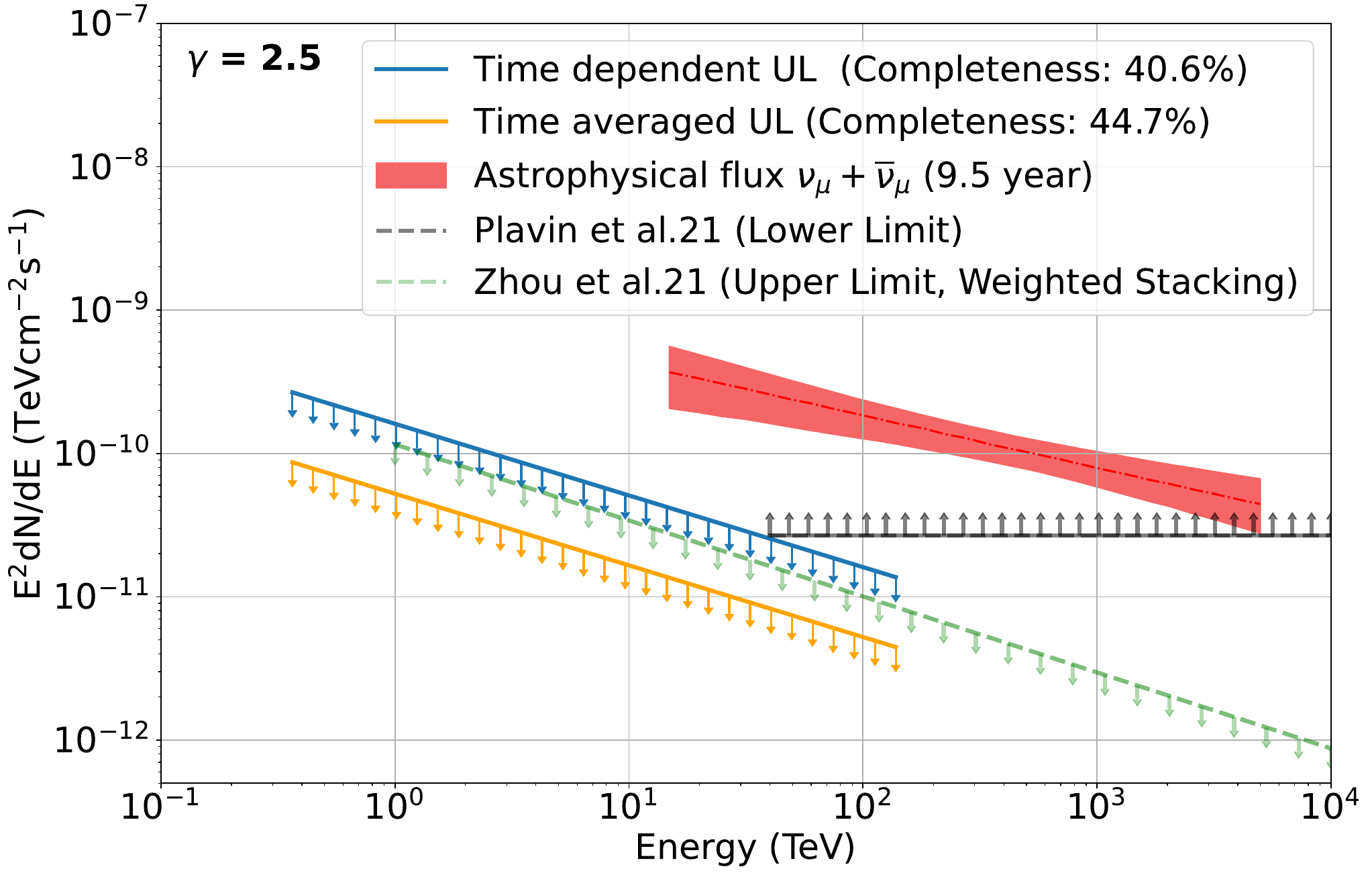}
   \end{tabular}
   \end{center}
  \caption{Upper Limits (UL) per neutrino ($\nu+\bar{\nu}$) flavor, for an index of 2.0 (left) and 2.5 (right solid line), derived from the time-dependent (blue) and time-averaged (orange solid line) analyses are shown here along with the lower limits (grey-dashed line) reported by \cite{Plavin_2021}. Note that while the samples and methodology used by the \cite{Plavin_2021}, \cite{zhou2021neutrino_agn} (green dashed line) and this work are different, making a 1:1 comparison difficult, they are shown here to highlight the differences in the results. The diffuse astrophysical muon neutrino flux measurements are taken from \cite{Diffuse_flux_9.5}. The energy range of the upper limits shown for the time-averaged analyses reflects the region where 90\% of detected signal neutrinos would fall. The energy range for the time-dependent scenario is kept similar to the time-averaged case.
    \label{fig:ul_2.0_blazar} }
\end{figure*}

Using a spectral index of 2.0 and 2.5, the energy-integrated upper limits for the two analyses at 100 TeV are given in Fig.~\ref{fig:ul_2.0_blazar}.
Both of these limits are shown after including the completeness correction described in Appendix\,A.  We also show the astrophysical diffuse flux reported by \cite{Diffuse_flux_9.5} and the lower limits given by \cite{Plavin_2021}. The energy range for the upper limits in the figure depicts the region where 90\% of the signal neutrinos with this spectrum will be detected. 
We calculate the 90\% sensitivity for both scenarios by determining the mean 90\% confidence level upper limit expected in the absence of signal \citep{ul_method_neutrino_detec_hill_2002}, calculated both in terms of flux and the number of neutrino events. The sensitivity of both the scenarios (in terms of $E^2dN/dE$ flux) is comparable within uncertainties, with a value of $\sim2\times10^{-12}$\,TeV\,cm$^{-2}$s$^{-1}$ for a spectral index of 2.0, with the time-dependent sensitivity being slightly higher. The discovery potential, defined as the signal strength leading to $5\sigma$ deviation for 50\% of all cases, is found to be $\sim1\times10^{-11}$\,TeV\,cm$^{-2}$s$^{-1}$ (time dependent) and $\sim8\times10^{-12}$\,TeV\,cm$^{-2}$s$^{-1}$ (time averaged), in terms of $E^2dN/dE$ flux for a spectral index of 2.0. However, the statistical significance of the time-dependent analysis, used to derive the upper limit, is also higher for the time-dependent analysis, giving it a higher upper limit (See also Fig~\ref{fig:background_dist} in Appendix B which shows the median and best-fit TS value used to calculate the sensitivity and upper limit respectively).

\begin{figure*}
\centering
\includegraphics[width=.6\textwidth,trim={0.0cm 0.1cm 0.0cm 0.0cm}]{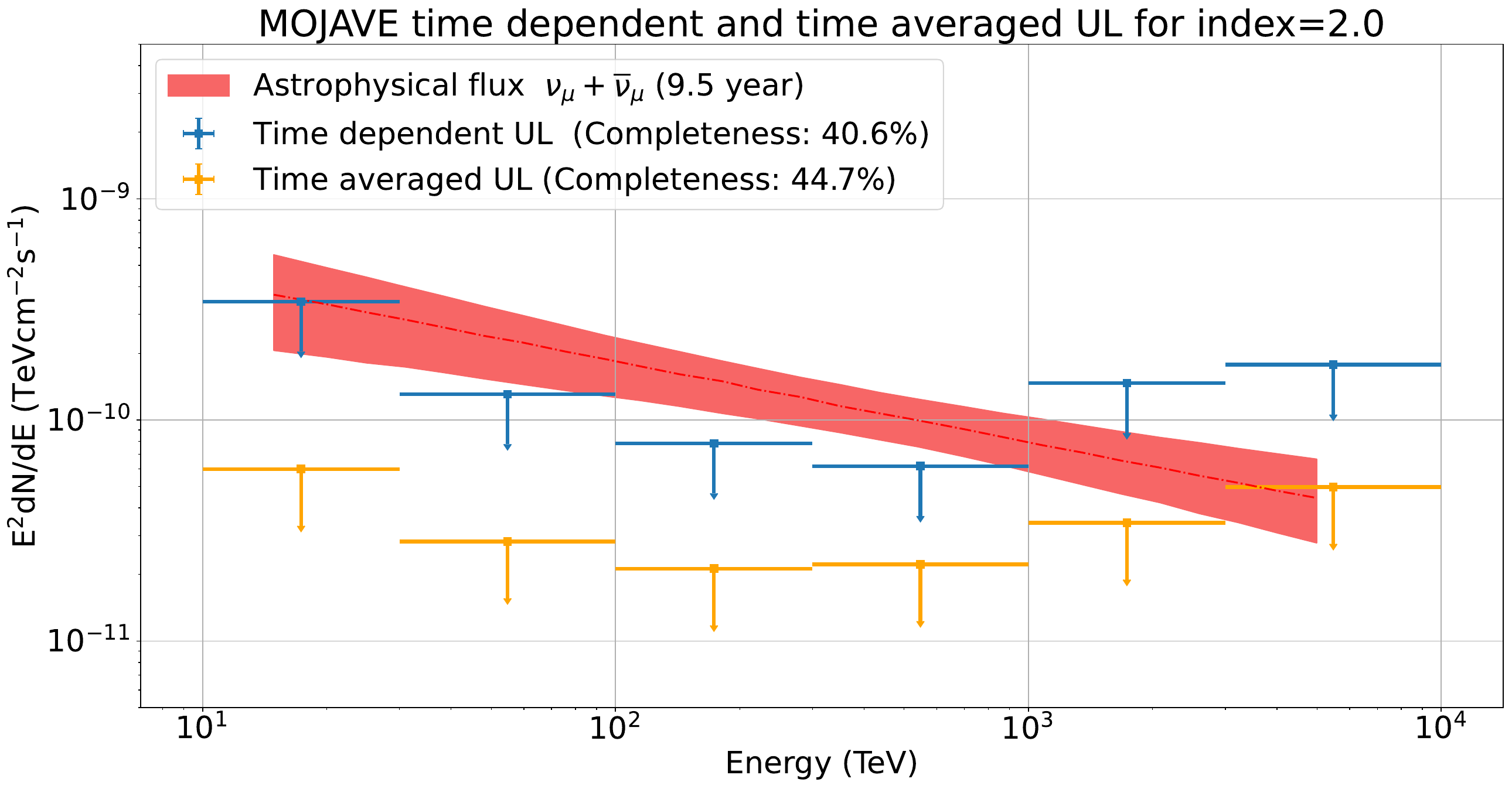}
\caption{Differential Upper Limits (UL), for an index of 2.0, derived from the time-dependent (blue) and time-averaged (orange) analyses in half-decade energy bins are shown here. The diffuse astrophysical muon neutrino flux measurements are taken from \cite{Diffuse_flux_9.5}. Note that the upper limits shown here are differential and binned with energy, while the astrophysical flux is an energy-integrated measurement shown here only for reference.
}
\label{fig:diff_ul_2.0_blazar}
\end{figure*}

We also report the differential upper limits for different energy bins for both of our analyses, which is shown in Figure.~\ref{fig:diff_ul_2.0_blazar}. The differential upper limit highlights the energy range where the study is most sensitive. The differential upper limits reduce the dependence on spectral assumptions and give a per energy bin upper limit. These estimates can also be used to highlight the energies at which neutrino production in AGN may be suppressed due to photon intensities or relevant production mechanisms, which can be used for AGN modeling studies. 

\subsection{Comparison with other studies:}
\label{comparision_with_previous_studies}
Previous studies have worked on similar analyses using VLBA radio data from AGN to search for correlations and reported limits on the total neutrino flux from these AGN \citep{Plavin_2020a,Plavin_2021,zhou2021neutrino_agn,christina_alert_plavin_fermi}. 

The lower limit reported by \cite{Plavin_2021}, also shown in Fig.~\ref{fig:ul_2.0_blazar}, lies above the upper limits provided here, even after the inclusion of a completeness correction. While adding temporal information for the time-dependent analysis changes the upper limits with respect to the time-averaged limit, it still lies below the \cite{Plavin_2021} results, ruling out the reported lower limits. 
Note that the radio and neutrino datasets and analysis methodology used for the two studies are different, which makes a direct comparison between the two studies difficult. This work uses a stacking approach with time-dependent MOJAVE radio data while the \cite{Plavin_2021} work makes use of the direction of the radio sources given in the RFC catalog to test for correlation with IceCube alerts.
However, a more detailed study using the complete IceCube alert dataset including additional information like signalness, was performed recently by \cite{christina_alert_plavin_fermi} which provides a direct comparison to the \cite{Plavin_2020a,Plavin_2021} results. 
In contrast to \cite{Plavin_2020a,Plavin_2021}, we find in  \cite{christina_alert_plavin_fermi} 
that the signal TS is compatible with the background and the significance goes down when a more sophisticated description of the spatial PDF is used as opposed to a simplified signal PDF modeled as a uniform distribution inside of the error contour.

On the other hand, \cite{zhou2021neutrino_agn} use the same neutrino dataset \citep[i.e.][]{icecube_10yr_data} as this work while using a more extensive radio sample (more stacked sources) instead of a completeness correction. 
However, when comparing the upper limits, the limits reported by \cite{zhou2021neutrino_agn} lie above the limits derived by this work. 
The change in sensitivity of this work as compared to \cite{zhou2021neutrino_agn} is mainly due to the inclusion of an energy PDF in the likelihood framework as described above, thereby reducing the upper limit.

%
%

\section{Conclusion:}
\label{sec:conclusion}

This work focused on using the AGN data published in the MOJAVE XV catalog to search for radio flux density-correlated neutrino emission using time-averaged and time-dependent analyses. We performed a stacking analysis and reported upper limits for both analyses as no significant correlation is found. When compared to the IceCube diffuse flux \citep{Diffuse_flux_9.5}, at 100 TeV and for a spectral index of 2.5, the upper limits derived are $\sim3\%$ and $\sim9\%$ for the time-averaged and time-dependent case. Note that, as the spectral index of the diffuse flux is different from 2.5, the percentage comparison is done using the upper limit estimates at 100\,TeV.
We also compared the upper limits presented in this work with the reported limits of \cite{Plavin_2021} and \cite{zhou2021neutrino_agn}. 
While the study presented here has the limitation of using fewer radio sources, which is made up by using a completeness correction, it also has the advantage of using more neutrino information in a time-dependent stacking study. 

Based on the obtained results, this study shows that including time-dependent information in the form of light curves improves the statistical power of the stacking analysis, if the neutrino flux is directly correlated to the change in the radio flux. While the sensitivity for both analyses is similar (see Sec.~\ref{sec:results}), the time-dependent study includes the addition of temporal information which increases the best-fit TS values and changes the results. 
However, the time-dependent analysis depends on the light curves used to satisfy the model assumptions for this study. The MOJAVE XV dataset used here has per-source observations reported with a varying cadence with a few sources being observed only a couple of times. 
Additionally, as compared to the number of sources observed by VLBA, the number of MOJAVE XV catalog sources with time-dependent observations is limited and focused on a blazar-dominated sample with a few non-blazar AGN.
This can be improved upon, in the future, by making use of more AGN sources with observations performed with a good cadence, preferably from a monitoring campaign.

%
%
\section{Acknowledgements}
\label{sec:Acknowledgements}
The IceCube collaboration acknowledges the significant contributions to this manuscript from Abhishek Desai and Justin Vandenbroucke. The authors would like to thank M Lister, Y. Kovalev and A. Plavin for useful discussions regarding the MOJAVE and RFC catalogs and required completeness calculation. The authors would also like to thank M.J. Romfoe for the useful comments on the paper draft.
The authors gratefully acknowledge the support from the following agencies and institutions:
USA {\textendash} U.S. National Science Foundation-Office of Polar Programs,
U.S. National Science Foundation-Physics Division,
U.S. National Science Foundation-EPSCoR,
U.S. National Science Foundation-Office of Advanced Cyberinfrastructure,
Wisconsin Alumni Research Foundation,
Center for High Throughput Computing (CHTC) at the University of Wisconsin{\textendash}Madison,
Open Science Grid (OSG),
Partnership to Advance Throughput Computing (PATh),
Advanced Cyberinfrastructure Coordination Ecosystem: Services {\&} Support (ACCESS),
Frontera computing project at the Texas Advanced Computing Center,
U.S. Department of Energy-National Energy Research Scientific Computing Center,
Particle astrophysics research computing center at the University of Maryland,
Institute for Cyber-Enabled Research at Michigan State University,
Astroparticle physics computational facility at Marquette University,
NVIDIA Corporation,
and Google Cloud Platform;
Belgium {\textendash} Funds for Scientific Research (FRS-FNRS and FWO),
FWO Odysseus and Big Science programmes,
and Belgian Federal Science Policy Office (Belspo);
Germany {\textendash} Bundesministerium f{\"u}r Bildung und Forschung (BMBF),
Deutsche Forschungsgemeinschaft (DFG),
Helmholtz Alliance for Astroparticle Physics (HAP),
Initiative and Networking Fund of the Helmholtz Association,
Deutsches Elektronen Synchrotron (DESY),
and High Performance Computing cluster of the RWTH Aachen;
Sweden {\textendash} Swedish Research Council,
Swedish Polar Research Secretariat,
Swedish National Infrastructure for Computing (SNIC),
and Knut and Alice Wallenberg Foundation;
European Union {\textendash} EGI Advanced Computing for research;
Australia {\textendash} Australian Research Council;
Canada {\textendash} Natural Sciences and Engineering Research Council of Canada,
Calcul Qu{\'e}bec, Compute Ontario, Canada Foundation for Innovation, WestGrid, and Digital Research Alliance of Canada;
Denmark {\textendash} Villum Fonden, Carlsberg Foundation, and European Commission;
New Zealand {\textendash} Marsden Fund;
Japan {\textendash} Japan Society for Promotion of Science (JSPS)
and Institute for Global Prominent Research (IGPR) of Chiba University;
Korea {\textendash} National Research Foundation of Korea (NRF);
Switzerland {\textendash} Swiss National Science Foundation (SNSF).

\bibliographystyle{aasjournal}
\bibliography{references}

%
%
\section*{Appendix A: Completeness Correction}

\begin{figure}
\centering
\includegraphics[width=.5\textwidth,trim={0.0cm 1.8cm 0.0cm 0.0cm}]{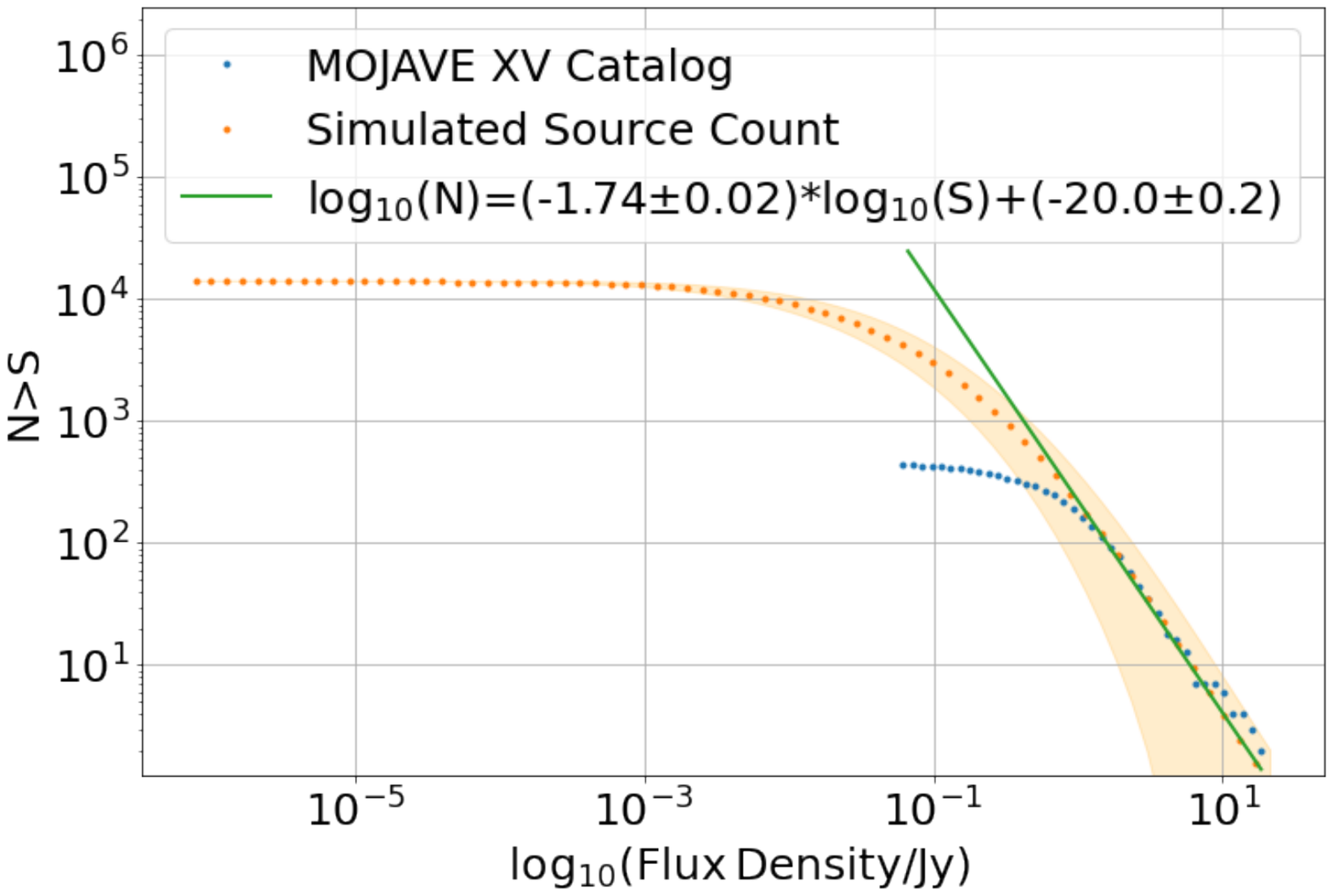}
\caption{Simulated source count distribution of the blazar-dominated sample (orange data points) as compared to the source count distribution of the MOJAVE XV sources (blue data points). The shaded region shows the one sigma error to the distribution due to varying the Lorentz factor and viewing angle parameters of the jets. The green line shows the fit at flux densities higher than $10^{-13}$erg sec$^{-1}$cm$^{-2}$ to the simulated sample.
}
\label{fig:completeness}
\end{figure}

While the MOJAVE sample is considered to be complete for flux densities greater than 1.5\,Jy, it is still considered to be a flux-limited and spatially-limited sample because of its declination constraints (limited by the region in the sky observed by VLBA telescopes; see Fig.~\ref{fig:aitoff}) and flux density measurement limitations of the telescopes used \citep{MOJAVEXV}. To account for these limitations, we compute the completeness correction for the sample by estimating the source count distribution of the sample. The source count distribution is the cumulative distribution of the number of sources greater than a particular flux (see Fig.~\ref{fig:completeness}). This is derived using the results of the population study performed by \cite{MOJAVEXVII} and reported as the MOJAVE XVII study. This involves modeling the flux ($\Phi$), luminosity ($L$) and redshift ($z$) relation using luminosity evolution parameterization similar to the ones used by \cite{Ajello_luminsity_function_fsrq}. To model the MOJAVE sources, we then use the parameters given by Model A of \cite{MOJAVEXVII} where $ \gamma=-3.1,k=8.0,\eta=-0.35$ and $\alpha=0$ for the equations:
\begin{equation}
    \Phi(L,z)\propto \Phi(L/e(z))
\end{equation}
\begin{equation}
    e(z)= (1+z)^k e^{z/\eta}
\end{equation}
\begin{equation}
    \Phi(L/e(z=0))\propto L^{\gamma}
\end{equation}

The cosmology parameters used in the calculation is taken from \cite{cosmology_planck_16} as $\Omega_0$=$0.308$, $\Omega_{\lambda}$=$0.692$ and $h_0$=$0.678$. In addition to these parameters, we also require the Lorentz factor distribution given by $N(\Gamma)d\Gamma\propto\Gamma^b$ where $b=-1.40$ and the viewing angle distribution given by $p(\theta)d\theta=\sin\theta$. To derive the simulated source count, the two PDFs are used to sample different values of $\Gamma$ and $\theta$ which are then used in the equations given above to derive the luminosity function. Once the luminosity function is derived, it is used to calculate the flux distribution and in turn the source count distribution. This procedure is repeated multiple times to get the uncertainties to the source count distribution which are shown as the shaded band in Fig.~\ref{fig:completeness}. Note that the mean source count distribution curve resulting from multiple simulations follows the  $\log_{10}N=(-1.63\pm0.02)\log_{10}S+(-19.8\pm0.03)$ equation for higher flux values where $N$ is the number of sources having a flux density greater than $S$. This is close to the accepted $\log N - \log S$ relation where $\log_{10}N=(-1.5)\log_{10}S$.
We also compare this simulated source count distribution to the sources given in the radio fundamental catalog which is complete for flux density measurements $>=150$\,mJy, and our simulation agrees with the observations for those flux density measurements.

Once the simulated source count distribution is derived, the area under the curve for the distribution is compared with the area under the curve for the observed MOJAVE source population. This gives the completeness correction for the population which takes into account both the flux and spatial limitations of the sample. This method adds dimmer sources to the population which might have been missed by the MOJAVE XV catalog. As the MOJAVE XV dataset and the population study done by \cite{MOJAVEXVII} is for a blazar-dominated source sample, the completeness derived using this method is similarly also for a blazar-dominated sample.

\section*{Appendix B: Background Distribution}

The TS distribution for a background-only case (${B}_i$) is shown here. This is derived by setting $n_s$ to 0 in Eq.~\ref{eq:likelihood} and running multiple standalone trials on the scrambled data which is derived by scrambling the R.A. of each event per trial. The shaded lines show the TS required for a $2\sigma$ and $3\sigma$ detection. The blue dashed line shows the best-fit TS values, which are used to derive the reported upper-limit measurement.

\begin{figure*}[ht!]
   \begin{center}
   \begin{tabular}{c}
    \includegraphics[width=.49\textwidth,trim={0.0cm 1.8cm 0.0cm 0.0cm}]{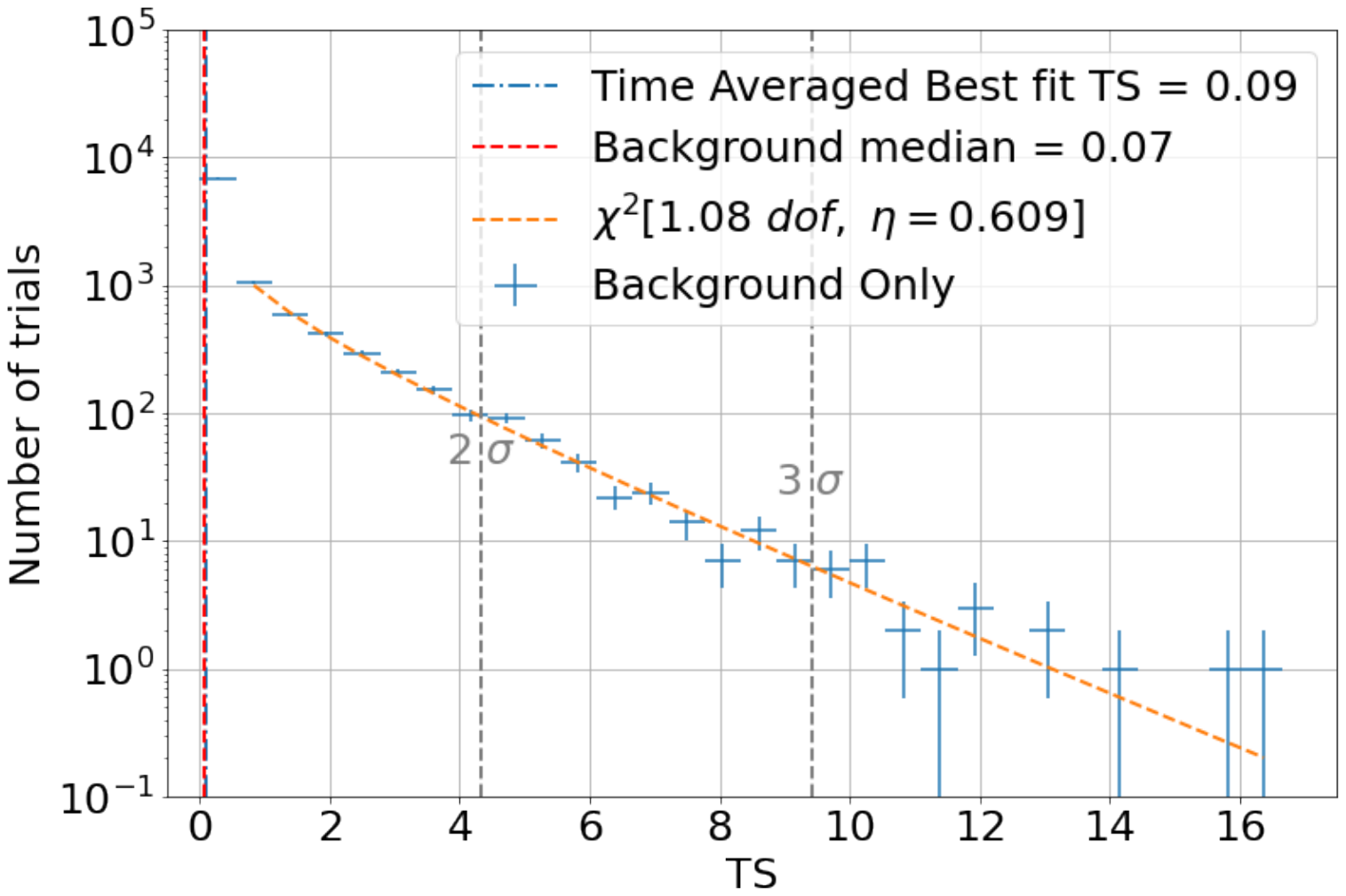}
    \includegraphics[width=.49\textwidth,trim={0.0cm 1.8cm 0.0cm 0.0cm}]{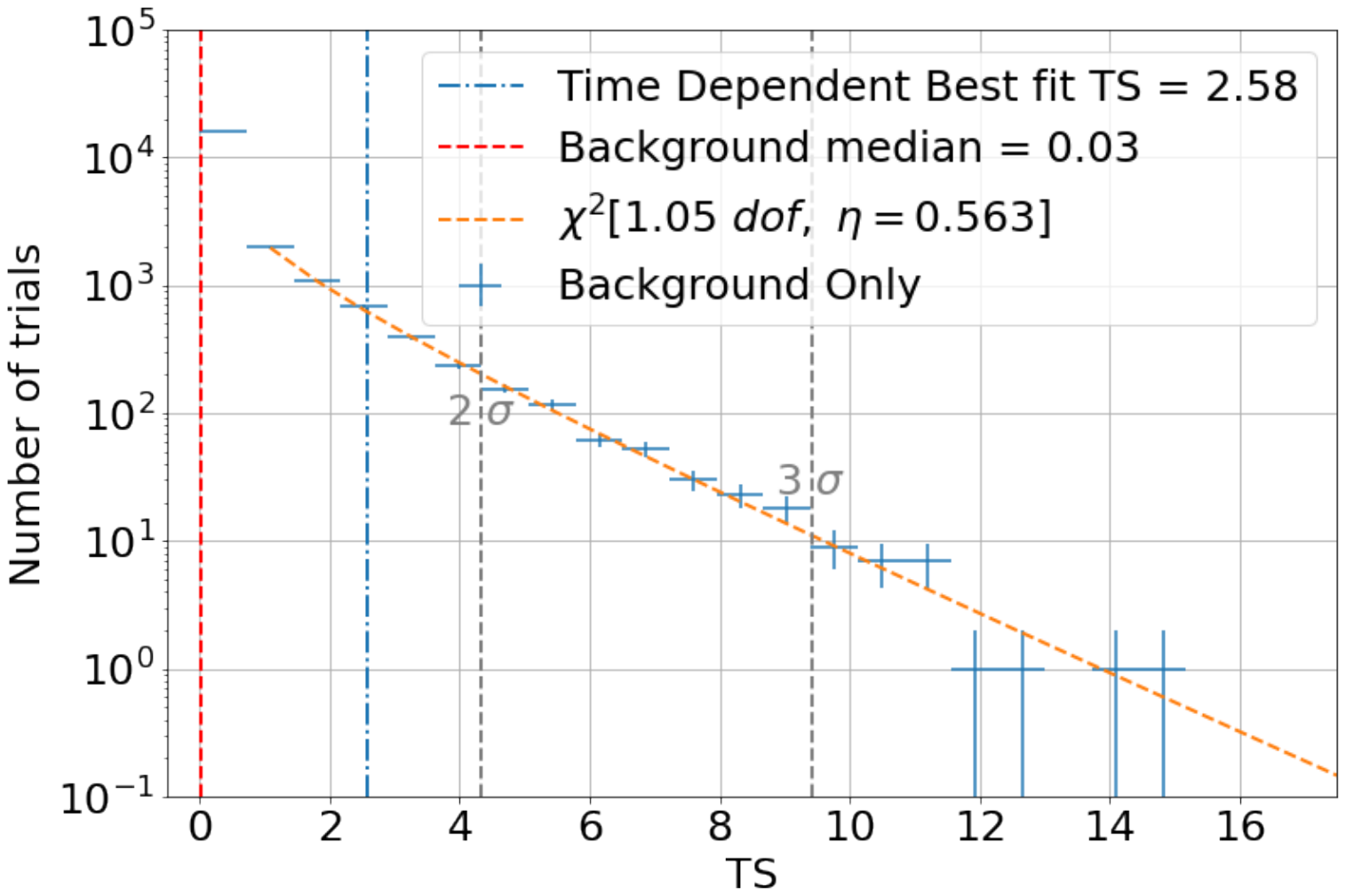}
   \end{tabular}
   \end{center}
  \caption{Background distribution plots derived for the time-averaged (left) and time-dependent (right) cases.}
  \label{fig:background_dist}
\end{figure*}

\end{document}